\documentclass[11pt,notitlepage]{article}
\usepackage{amssymb}
\usepackage{psfig}
\usepackage[dvips]{graphics}
\usepackage{graphicx}
\usepackage{subeqnarray}
\usepackage{subfigure}        

\newcommand{\bea}{\begin{eqnarray}}
\newcommand{\eea}{\end{eqnarray}}

\begin{document}
 \title{\bf SOLITON RESONANCES FOR MKP-II}

\author{ Jyh-Hao Lee$^{*}$ and Oktay K. Pashaev$^{\dagger}$\\
$^{*}$ Institute of Mathematics, Academia Sinica, Taipei, Taiwan \\
$^{\dagger}$ Department of Mathematics,
Izmir Institute of Technology, \\
 Urla-Izmir, 35430, Turkey\\}

\maketitle

\begin{abstract}

Using the second flow - the Derivative Reaction-Diffusion system,
and the third one of the dissipative SL(2,R) Kaup-Newell
hierarchy, we show that the product of two functions, satisfying
those systems is a solution of the modified
Kadomtsev-Petviashvili equation in 2+1 dimension with negative
dispersion (MKP-II). We construct Hirota's bilinear
representation for both flows and combine them together as the
bilinear system for MKP-II. Using this bilinear form we find one
and two soliton solutions for the MKP-II. For special values of
parameters our solution shows resonance behaviour with creation
of four virtual solitons. Our approach allows one to interpret the
resonance soliton as a composite object of two dissipative
solitons in 1+1 dimensions.
\end{abstract}

\section{Introduction}

The gauge theoretical formulation of the low dimensional gravity
models, like the Jackiw-Teitelboim model \cite{JT}, is based on
the Cartan-Einstein vielbein or the moving frame method. In terms
of these variables in 1+1 dimensions one deals with so called BF
gauge theory  and the zero curvature equations of motion,
providing a link with soliton equations \cite{MPS}. But in these
variables, as the "square root"
 of the pseudo-Riemannian metric, the soliton equations have dissipative
form, this is why we called them the dissipative solitons or
dissipatons \cite{MPS}. The dissipative version of the Nonlinear
Schrodinger equation with a rich resonance dynamics \cite{PL} is a
couple of nonlinear heat and anti-heat equations, which we called
the Reaction-Diffusion system (RD) \cite{MPS}, \cite{PL}. The
dissipaton of that system is the pair of two real functions, one
of which is exponentially growing, and another decaying, in space
and time. But the product of these two functions has the perfect
soliton form. As recently realized \cite{PF}, if dissipatons of
RD evolves with additional time variable according to the next
member after RD  of SL(2,R) AKNS hierarchy, with cubic
dispersion, then this product can be considered as a soliton of
2+1 dimensional Kadomtsev-Petviashvili equation with negative
dispersion (KP-II). This way the resonance behaviour of KP-II
solitons was found in terms of dissipatons of the 1+1 dimensional
models. Moreover, in this approach, the novel two resonance
soliton of KP-II with four virtual solitons was constructed and
interpreted as the degenerate four soliton solution \cite{PF}.
\par From another site, as was shown before \cite{LLP}, dissipative version of the
Derivative nonlinear Schrodinger equations (DNLS) also admits
dissipative soliton solutions with resonance interaction
\cite{ANZIAM}. Moreover, these resonances show the chirality
properties, propagating only in one direction. In the present
paper, following the strategy of paper \cite{PF}, by the recursion
operator of the Kaup-Newell (KN) hierarchy, first we construct
the next dissipative system of the hierarchy with cubic
dispersion. Then, using these two members of SL(2,R) Kaup-Newell
hierarchy, we show that the product of dissipaton functions is
solution of the modified Kadomtsev-Petviashvili equation with
negative dispersion (MKP-II) (Section 2). In Section 3
bilinearization of the two flows allows us to find bilinear form
for MKP-II. Chiral resonance dissipatons of DRD and their
geometrical meaning are considered in Section 4. One and two
soliton solutions of MKP-II are constructed in Section 5. We show
the resonance character of these solitons interaction and the
chirality property posing restriction on the soliton collision
angles. In conclusion we discuss main results of this paper.

\section{MKP-II from Kaup-Newell Hierarchy}

The KN hierarchy has the form \cite{Yan}
$$ \left(\matrix{q_{t_{n}} \cr r_{t_{n}}\cr} \right) = J L^n  \left(\matrix{q \cr r\cr}\right)
\eqno(2.1)$$ where the operator
$$ \left(\matrix{0 & \partial \cr \partial & 0\cr}\right)  \eqno(2.2)$$
is the first symplectic form, while
$$L = {1\over 2} \left(\matrix{-\partial - r \partial^{-1}q
\partial & -r\partial^{-1}r \partial \cr -q\partial^{-1}q\partial &
\partial - q\partial^{-1}r\partial\cr}\right)  \eqno(2.3)$$
is the recursion operator of the hierarchy, and $\partial \equiv
\partial/\partial x$. The second flow of the hierarchy is the
system
$$q_{t_{2}} = {1\over 2} [q_{xx} + (q^2 r)_x ],\eqno(2.4a)$$
$$r_{t_{2}} = {1\over 2} [-r_{xx} + (r^2 q)_x ],\eqno(2.4b)$$
while the third one is
$$q_{t_{3}} = -{1\over 4} [q_{xx} + 3 rq q_x + {3\over 2}(r^2 q^2)q ]_x,\eqno(2.5a)$$
$$r_{t_{3}} = -{1\over 4} [r_{xx} - 3 rq r_x + {3\over 2}(r^2 q^2)r ]_x,\eqno(2.5b)$$
For the SL(2,R) case of KN hierarchy we have real
time variables $t_2, t_3$ which we denote as $y \equiv t_2/2$, and $t \equiv - t_3/4$.
In this case functions
$q$ and $r$ are real,
and we denote them as
$$ e^+ \equiv q,\,\,\,\,e^{-} \equiv - r \eqno(2.6)$$
Then we have the DRD system \cite{LLP}
$$e^+_{y} =  e^+_{xx} - (e^+ e^- e^+)_x ,\eqno(2.7a)$$
$$e^-_{y} = -e^-_{xx} - (e^+ e^- e^-)_x ,\eqno(2.7b)$$
and
$$e^+_{t} =  e^+_{xxx} - 3(e^+ e^- e^+_x)_x + {3\over 2}((e^+e^-)^2e^+)_x ,\eqno(2.8a)$$
$$e^-_{t} =  e^-_{xxx} + 3(e^+ e^- e^-_x)_x + {3\over 2}((e^+e^-)^2e^-)_x ,\eqno(2.8b)$$
 \par
Now we consider the pair of functions of three variables
$e^+(x,y,t)$ and $e^-(x,y,t)$ satisfying the systems (2.7) and
(2.8). These systems are compatible since they belong to the same
hierarchy for different times. This can be also checked directly
from compatibility condition $e^{\pm}_{ty} = e^{\pm}_{yt}$ by
using following conservation laws for Eqs.(2.7) and (2.8)
respectively
$$(e^+e^-)_y = [(e^+_x e^- - e^+ e^-_x) - {3\over 2}(e^+e^-)^2]_x ,\eqno(2.9)$$
$$(e^+e^-)_t = [(e^+ e^-)_{xx} - 3(e^+_x e^-_x) + 3(e^+ e^-)(e^+ e^-_x - e^+_x e^-)
 + {5\over 2}(e^+e^-)^3]_x $$

{\bf Proposition 1}. Let the functions  $e^+(x,y,t)$ and
$e^-(x,y,t)$, are solutions of the systems (2.7) and (2.8)
simultaneously. Then, the function $U(x,y,t) \equiv e^+ e^-$
satisfies the modified Kadomtsev-Petviashvili equation (MKP-II)
$$(-4U_t + U_{xxx} - {3\over 2} U^2 U_x  - 3 U_x \partial^{-1}_x U_y)_x = -3 U_{yy}
\eqno(2.10)$$ or written in another form
 $$-4U_t + U_{xxx} - {3\over 2} U^2 U_x  - 3 U_x W = -3 W_{y} \eqno(2.11a)$$
$$W_x = U_y \eqno(2.11b)$$
(The second form appears from the first one by introducing
auxiliary variable $W$ according to Eq.(2.11b) and integration in
variable $x$).
\par The proof is straightforward. From definition of $U$ and Eqs(2.7),(2.9)
we have
$$U_y =  [(e^+_x e^- - e^+ e^-_x) - {3\over 2}U^2]_x ,\eqno(2.12)$$
$$U_{yy} = [U_{xxx} - 4(e^+_x e^-_x)_x - 3U_x (e^+_x e^- - e^+ e^-_x) -
U (e^+_x e^- - e^+ e^-_x)_x - {3\over 2}(U^2)_y]_x $$ From
another site Eqs.(2.8),(2.9) give
$$U_t = [U_{xx} - 3(e^+_x e^-_x) - 3 U(e^+_x e^- - e^+ e^-_x) + {5\over 2}U^3]_x \eqno(2.13)$$
$$U_{tx} = [U_{xx} - 3(e^+_x e^-_x) - 3 U(e^+_x e^- - e^+ e^-_x) + {5\over 2}U^3]_{xx}
\eqno(2.14)$$ First we combine Eqs.(2.12) and  (2.14) to cancel
term $e^+_x e^-_x$, then use Eq.(2.12) to exclude $e^+_x e^- -
e^+ e^-_x$ and its derivative, according to
$$ (e^+_x e^- - e^+ e^-_x)_x = U_y - {3\over 2}(U^2)_x \eqno(2.15)$$
and integrating once
$$(e^+_x e^- - e^+ e^-_x) = \partial^{-1}_x U_y - {3\over 2}U^2 \eqno(2.16)$$
As a result we arrive with the MKP-II (2.10)

\section{Bilinear Form for the Second and Third flows}

Now we will construct bilinear representation for systems (2.7)
and (2.8) to find solutions of MKP-II according to our Proposition
1. In our paper \cite{LLP} we applied the Hirota bilinear method
to integrate RDR (2.7). Now we will apply the same method to
Eq.(2.8) and MKP-II. As was noticed in \cite{LLP}, the standard
Hirota substitution as the ratio of two functions doesn't work
directly for $e^+$ and $e^-$. (This fact also is related with
complicated analytical structure of DNLS \cite{L}). To have the
standard Hirota substitution, following \cite{LLP} we first
rewrite the systems (2.7) and (2.8) in terms of new functions
$Q^+$, $Q^-$:
$$e^+ = e^{+\int^x Q^+ Q^-} Q^+,\,\,\,\,  e^- = e^{-\int^x Q^+ Q^-} Q^-,\eqno(3.1)$$
and as result we have the systems
$$Q^+_{y} =  Q^+_{xx} + Q^+ Q^+ Q^-_x - {1\over 2} (Q^+ Q^-)^2 Q^+,\eqno(3.2a)$$
$$Q^-_{y} = -Q^-_{xx} + Q^- Q^- Q^+_x + {1\over 2} (Q^+ Q^-)^2 Q^- ,\eqno(3.2b)$$
and
$$Q^+_{t} =  Q^+_{xxx} + 3Q^+_x Q^-_x Q^+ - {3\over 2}(Q^+ Q^-)^2 Q^+_x ,\eqno(3.3a)$$
$$Q^-_{t} =  Q^-_{xxx} - 3Q^+_x Q^-_x Q^- - {3\over 2}(Q^+ Q^-)^2 Q^-_x ,\eqno(3.3b)$$
 \par
Then, due to the fact that
$$Q^+ Q^- = e^+ e^- = U ,\eqno(3.4)$$
the systems (3.2), (3.3) provide also solution of MKP-II which we
can formulate as below.

\par
{\bf Proposition 2}.  Let the functions  $Q^+(x,y,t)$ and
$Q^-(x,y,t)$, are solutions of the systems (3.2) and (3.3)
simultaneously. Then, the function $U(x,y,t) \equiv Q^+ Q^-$
satisfies the modified Kadomtsev-Petviashvili equation (MKP-II)
(2.10) or (2.11).

\par
To solve the systems (3.2)  and (3.3) we introduce four real
functions $g^+, g^-, f^+, f^-$ according to the formulas
$$Q^+ = {g^+\over f^+},\,\,\,\,\,\,\,\,\,Q^- = {g^-\over f^-},\eqno(3.5)$$
or  using Eqs.(3.1) and (3.4) for the original variables $e^+$ and
$e^-$ we have the following substitution
$$
e^{+} = {g^{+} f^{+} \over (f^{-})^2},\,\,\,\,\ e^{-} = {g^{-}
f^{-} \over (f^{+})^2}. \eqno(3.6)$$
Then the system (3.2)
bilinearizes in the form
$$(D_y \mp D^2_x)(g^{\pm}\cdot f^{\pm}) = 0, \eqno(3.7a)$$
$$D^2_x (f^+\cdot f^-) + {1\over 2} D_x (g^+\cdot g^-) = 0, \eqno(3.7b)$$
$$D_x(f^+ \cdot f^-) - {1\over 2} g^+ g^- = 0.\eqno(3.7c)$$
In a similar way, for the system (3.3) we have the next bilinear
form
$$(D_t - D^3_x)(g^{\pm}\cdot f^{\pm}) = 0, \eqno(3.8a)$$
$$D^2_x (f^+\cdot f^-) + {1\over 2} D_x (g^+\cdot g^-) = 0, \eqno(3.8b)$$
$$D_x(f^+ \cdot f^-) - {1\over 2} g^+ g^- = 0.\eqno(3.8c)$$
Comparing these two bilinear forms we can see that the second and
the third equations in both systems (3.7),(3.8) are of the same
form. This is why for simultaneous solution of both Eqs.(3.2),
(3.3) we have the next bilinear system
$$(D_y \mp D^2_x)(g^{\pm}\cdot f^{\pm}) = 0, \eqno(3.9a)$$
$$(D_t - D^3_x)(g^{\pm}\cdot f^{\pm}) = 0, \eqno(3.9b)$$
$$D^2_x (f^+\cdot f^-) + {1\over 2} D_x (g^+\cdot g^-) = 0, \eqno(3.9c)$$
$$D_x(f^+ \cdot f^-) - {1\over 2} g^+ g^- = 0.\eqno(3.9d)$$
>From the last equation we have
$$U = e^+ e^- = Q^+ Q^- = {g^+ g^-\over f^+ f^-} = 2 {D_x (f^+\cdot f^-)\over f^+ f^-}
= 2 {f^+_x f^- - f^+ f^-_x\over f^+ f^-} $$ which provides for
solution of MKP-II the following formula
$$U = 2(\ln {f^+\over f^-})_x \eqno(3.10)$$

\section{Resonance Solitons of DRD}
\par
First of all we consider DRD (2.7)
 as an evolution equation where $y = x_0 = t$ is interpreted as
 the time variable.
\subsection{Chiral Dissipative Soliton Solution}\label{subsec:1d}
For one dissipaton solution we have $$ g^{\pm} = e^{\eta^{\pm}},
\,\,\,\, f^{\pm} = 1 + e^{\phi^{\pm}} e^{\eta^{+} +
\eta^{-}},\,\,\, e^{\phi^{\pm}} = \pm {k^{\mp} \over 2 (k^{+} +
k^{-})^2},\eqno(4.1)$$ where $\eta^{\pm} = k^{\pm}x \pm
(k^{\pm})^2 t + \eta^{\pm}_{0}$. The regularity requires that we
choose conditions $k^{-} > 0$ and $k^{+} < 0$ so that we have
dissipaton $$ Q^{\pm} = {e^{\pm (\eta^{+} - \eta^{-})/2 -
\alpha_{\pm}/2} \over 2 \cosh {1 \over 2}(\eta^{+} + \eta^{-} +
\alpha_{\pm}) }, \eqno(4.2)$$ with solitonic density $$ e^{+}e^{-}
= Q^{+}Q^{-} = {2 k^2 \over \sqrt {v^2 - k^2}\cosh k(x - vt -
x_{0}) + v}. \eqno(4.3)$$ In the last equation we introduced the
dissipaton's amplitude and velocity, $k = k^{+} + k^{-}$ and $v =
k^{-} - k^{+}$ correspondingly, in terms of which the above
conditions mean that velocity of dissipaton is bounded from below
by $k$. It is worth noting that in contrast with dissipatons of RD
system \cite{PL}, in our case no critical value from above for
dissipaton's velocity exists.

For the mass and momentum densities we have
$$
\rho = e^{+} e^{-}  = 2 \partial_{x} \ln {f^{+} \over f^{-}},
\eqno(4.4)$$
$$
p =  {1 \over 2}( e^{+} \partial_{x} e^{-} - e^{-} \partial e^{+}
+ (e^{+} e^{-})^2)
 =  \partial^2_{x} \ln f^{+} f^{-},
\eqno(4.5)$$ which allows one calculate the corresponding
conserved quantities $$ M = \int^{+\infty}_{-\infty} \rho dx = 2
\ln {f^{+} \over f^{-}} |^{+\infty}_{-\infty},\,\,\, P =
\int^{+\infty}_{-\infty} p dx = \partial_{x} \ln f^{+} f^{-}
|^{+\infty}_{-\infty}. $$ Then for mass and momentum of a single
dissipaton we obtain $$ M = \ln ({v + k \over v -
k})^2,\,\,\,\,\,\, P = k. \eqno(4.7)$$ Due to relation $|v|
> |k|$ the mass $M$ is positive. Rewriting momentum in the
canonical form $ k = \mu v$, we find for the effective mass $\mu
= \tanh M/4$.

\subsection{Geometrical Interpretation}
The model has geometrical interpretation of two dimensional
pseudo-Riemannian spacetime with constant scalar curvature $R =
\Lambda < 0$ \cite{MPS},\cite{PL}. This model is known as the
Jackiw-Teitelboim gravity \cite{JT}. It admits gauge theoretical
formulation as the BF theory. The  gauge potentials are
Cartan-Einstein zweibein fields $e^{\pm}_{\mu}$, so that the
metric is $g_{\mu\nu} = (e^{+}_{\mu}e^{-}_{\nu} +
e^{+}_{\nu}e^{-}_{\mu})/2$,
 and the spin-connection $\omega_{\mu}$,
($\mu = 0,1$). Then equations of motion
$$
D^{\mp}_{\mu}e^{\pm}_{\nu} = D^{\mp}_{\nu}e^{\pm}_{\mu},
\eqno(4.8)$$

$$
\partial_{\mu} \omega_{\nu} - \partial_{\nu} \omega_{\mu}
= -{\Lambda \over 4}(e^{+}_{\mu}e^{-}_{\nu} -
e^{+}_{\nu}e^{-}_{\mu}), \eqno(4.9)$$

where $D^{\pm}_{\mu} =
\partial_{\mu} \pm \omega_{\mu}$, have meaning of torsionless and
the constant curvature conditions respectively. We fix the gauge
freedom and the corresponding evolution by the following
conditions on Lagrange multipliers $$ e^{\pm}_{0} = \pm
(\partial_{1} \mp {\Lambda \over 4} \mp e^{+}e^{-})e^{\pm},
\eqno(4.10)$$ and spin connections $$ \omega_{0} = {\Lambda^2
\over 16} + {\Lambda \over 4}e^{+}e^{-}, \,\,\,\,\,\omega_{1} =
-{\Lambda \over 4}. \eqno(4.11)$$ Then after the identification
$(t,x) = (x_{0},x_{1})$ the system (4.8), (4.9) reduces to DRD
(2.7). It is worth noting that in contrast to the RD \cite{PL} in
our case the scalar curvature disappears from equations of motion
but is still present in the linear problem. Moreover it has the
meaning of the squared spectral parameter $\Lambda  = - 8
\lambda^2$.

The metric tensor component $g_{00}$ in terms of transformed
variables (3.1) is given by $$ g_{00} = e^{+}_{0}e^{-}_{0} =
-(\partial_{1} - {\Lambda \over 4})Q^{+} (\partial_{1} + {\Lambda
\over 4})Q^{-}. \eqno(4.12)$$ Calculating it for one dissipaton
(4.2) we find that event horizon  $g_{00} = 0$ exists only for the
bounded velocity $-k + (-\Lambda/2) < v < k + (-\Lambda/2)$ at
the distance $$ x_{H} - v t_{H} - x_{0H} = {1 \over k} \ln {4k^2
(k - v - {\Lambda \over 2}) \over (k + v) (k + v + {\Lambda \over
2})}. \eqno(4.13)$$ In contrast to the RD \cite{PL} dissipaton
with two symmetrical event horizons reflecting two directions of
motion, we have now only one directional motion and call the
corresponding single event horizon as the {\it chiral event
horizon}.

\section{Resonance Solitons of MKP-II}
\par
Now we consider a solution of the system (3.9), giving 2+1
dimensional solution of MKP-II. For one-soliton solution we have
$$g^{\pm} = e^{\eta^{\pm}_{1}},\,\,\,f^{\pm} = 1 + e^{\phi^{\pm}_{11}}e^{\eta^{+}_{1} + \eta^{-}_{1}},\,\,\,
e^{\phi^{\pm}_{11}} = \pm {k^{\mp}_{1} \over 2 (k^{+}_{1} +
k^{-}_{1})^2},\eqno(5.1)$$ where, $\eta^{\pm}_{1} = k^{\pm}_{1}x
\pm (k^{\pm}_{1})^2 y + (k^{\pm})^3 t + \eta^{\pm}_{0}$. The
regularity condition requires $k^{+}_{1} \leq 0$, $k^{-}_{1} \geq
0$. Then we have
$$U(x,y,t)  = {2k^2 \over \sqrt{p^2 - k^2}
\cosh {k(x - p y + {k^2 + 3p^2\over 4}t - a_{0})} + p},
\eqno(5.2)$$ where $k = k^{+}_{1} + k^{-}_{1}$, $p = k^{-}_{1} -
k^{+}_{1} > 0$, and bounded from the below parameter $p^2 > k^2$
is positive $p > 0$. The geometrical meaning of this parameter is
$p^{-1} = \tan \alpha $, where $\alpha $ is the slope of the
soliton line. Due to the condition $p > 0$, the direction of this
line is restricted between  $0 < \alpha < \pi/2 $. (This is the
space analog of the chirality property of dissipaton in 1+1
dimensions for DNLS [7], when it propagates only in one
direction.) The velocity of soliton is two dimensional vector
${\bf v} = (\omega, -\omega/p)$, where $\omega = (k^2 + 3p^2)/4$,
directed at angle $\gamma$ to the soliton line, where $\cos
\gamma = 1 - 1/p^2$. When $p = 1$, the velocity of soliton is
orthogonal to the soliton line.

\par
For two soliton solution we have

$$g^{\pm} = e^{\eta^{\pm}_{1}} + e^{\eta^{\pm}_{2}} +
\alpha^{\pm}_{1}e^{\eta^{+}_{2} + \eta^{-}_{2} + \eta^{\pm}_{1}} +
\alpha^{\pm}_{2}e^{\eta^{+}_{1} + \eta^{-}_{1} + \eta^{\pm}_{2}},$$

$$f^{\pm} = 1 + \sum_{i,j = 1}^2 e^{\phi^{\pm}_{ij}} e^{\eta^{+}_{i} + \eta^{-}_{j}}
+ \beta^{\pm}e^{\eta^{+}_{1} + \eta^{-}_{1} + \eta^{+}_{2} + \eta^{-}_{2}},$$

where $\eta^{\pm}_{i} = k^{\pm}_{i}x \pm (k^{\pm}_{i})^2 y +
(k^{\pm})^3 t + \eta^{\pm}_{i0}$, $k^{nm}_{ij} \equiv (k^{n}_{i} + k^{m}_{j})$ and

$$\alpha^{\pm}_{1} = \pm {1 \over 2}{k^{\mp}_{2} (k^{\pm}_{1} -
k^{\pm}_{2})^2\over (k^{+-}_{22})^2(k^{\pm\mp}_{12})^2 },\,\,\,
\alpha^{\pm}_{2} = \pm {1 \over 2}{k^{\mp}_{1} (k^{\pm}_{1} -
k^{\pm}_{2})^2\over (k^{+-}_{11})^2(k^{\pm\mp}_{21})^2 },$$
$$\beta^{\pm} = {(k^{+}_{1} - k^{+}_{2})^2 (k^{-}_{1} -
k^{-}_{2})^2 \over 4(k^{+-}_{11}k^{+-}_{12}k^{+-}_{21}k^{+-}_{22})^2} k^{\mp}_{1}k^{\mp}_{2},$$

$$e^{\phi^{\pm}_{ii}} = \pm {k^{\mp}_{i} \over 2 (k^{+-}_{ii})^2},\,\,\,
e^{\phi^{+}_{ij}} = {k^{-}_{j} \over 2 (k^{+-}_{ij})^2}, \,\,\,
e^{\phi^{-}_{ij}} = - {k^{+}_{i}  \over 2 (k^{+-}_{ij})^2}.$$ The
regularity conditions now are the same as for one soliton
$k^{+}_{i} \leq 0$, $k^{-}_{i} \geq 0$. Then this solution
describes a collision of two solitons propagating in plane and at
some value of parameters creating the resonance states (Fig.1a,b).

\begin{figure}[ht]
\begin{center}
\mbox{
\subfigure[Fig.1a]{\includegraphics[height= 6.4cm]{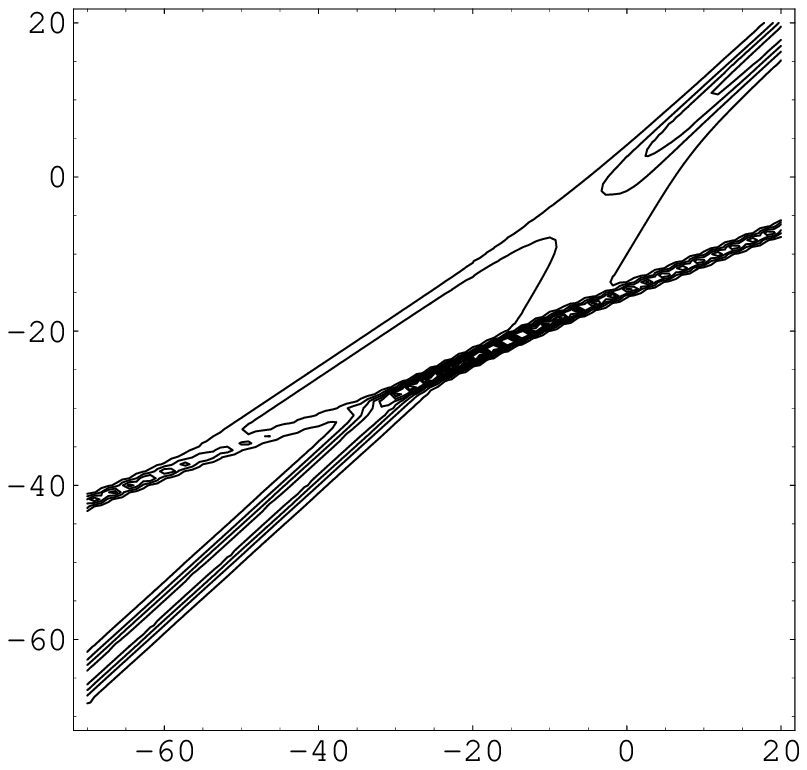}}
\subfigure[Fig.1b]{\includegraphics[height= 6.4cm]{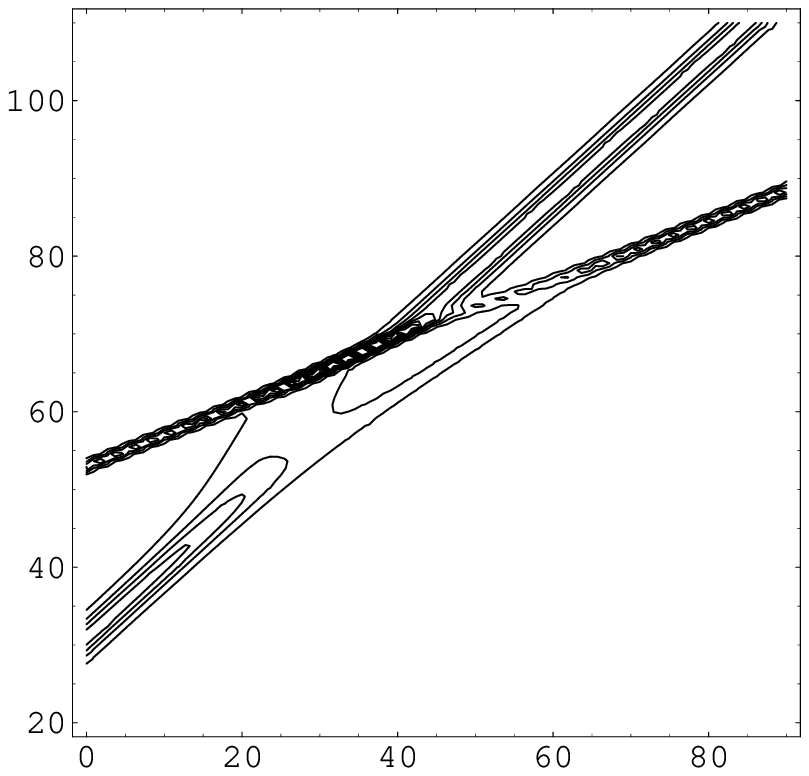}}
 }
\end{center}
\end{figure}

\section{Conclusion}

In the present paper we have constructed virtual soliton
resonance solutions for MKP-II in terms of dissipative solitons of
1+1 dimensional equations as the Derivative Reaction-Diffusion
equation and its higher member of SL(2,R) Kaup-Newell hierarchy.
The difference with the KP-II resonance is in the additional
restrictions on solitons angles from regularity conditions.

When this paper has been finished, on the Conference on Nonlinear
Physics in Gallipoli, Italy, June-July 2004, Konopelchenko
attracted our attention to the relations between MKP equation and
1+1 dimensional models by the symmetry reduction of 2+1
dimensional models \cite{KS}, \cite{ChengLi}. But in paper
\cite{KS} a relation of MKP only with Burgers hierarchy has been
established. While paper \cite{ChengLi} relates MKP with
derivative NLS in the Nakamura-Chen form but not in the
Kaup-Newell form. Moreover no results on resonance solitons in
those papers are found.

 {\bf Acknoweledgments}
\par\noindent
The authors would like to thanks B. Konopelchenko for useful
remarks and Y. Kodama for fruitful discussions. This work was
supported partially by Institute of Mathematics, Academia Sinica,
Taipei, Taiwan and Izmir Institute of Technology, Izmir, Turkey.

\end{document}